\documentclass[twocolumn,prd,nofootinbib,aps,prl,floats,floatfix,amsmath,amssymb,longbibliography,secnumarabic]{revtex4-1} %
\usepackage[english]{babel}

\usepackage[letterpaper,top=2cm,bottom=2cm,left=3cm,right=3cm,marginparwidth=1.75cm]{geometry}

\usepackage{amsmath}
\usepackage{graphicx}
\usepackage[colorlinks=true,citecolor=red]{hyperref}

\newcommand{\be}{\begin{equation}}
\newcommand{\ee}{\end{equation}}
\def\sfrac#1#2{{\textstyle{#1\over #2}}}
\newcommand{\bea}{\begin{eqnarray}}
\newcommand{\eea}{\end{eqnarray}}
\newcommand{\nn}{\nonumber}

\begin{document}

\title{Comment on ``Dark Energy from Time Crystals''}
\author{James M.\ Cline}
\email{jcline@physics.mcgill.ca}
\affiliation{McGill University Department of Physics \& Trottier Space Institute, 3600 Rue University, Montr\'eal, QC, H3A 2T8, Canada}

\begin{abstract}
It was recently proposed (\url{https://arxiv.org/pdf/2502.08887}) that the time crystal Lagrangian 
introduced by Shapere and Wilczek in 2012 could be a model of dark energy.  I point out that the model has an instability that drives its energy density to negative values, which may render it unsuitable as a model of dark energy.

\end{abstract}

\maketitle

Recently Ref.\ \cite{Mersini-Houghton:2025ybm} proposed that a higher-derivative scalar field theory, with a Lagrangian of the form
\be
    {\cal L} = -\kappa X + \lambda X^2 - V(\phi)\,,
    \label{lageq}
\ee
could be a viable model of dark energy, where $X = \sfrac12 \partial_\mu\phi\partial^\mu\phi = \sfrac12(\dot\phi^2 - (\nabla\phi)^2)$,\footnote{The author of Ref.\ \cite{Mersini-Houghton:2025ybm} wrote
$X = \sfrac12\dot\phi^2$ only, since she was focusing on time-independent solutions, but I will assume that the underlying Lagrangian is Lorentz invariant.}
and $\lambda,\kappa$ are positive constants.
This is similar to a model that was proposed in 
Ref.\ \cite{Shapere:2012nq} in the context of point particle mechanics, and dubbed ``time crystals", by virtue of spontaneously breaking time translation symmetry in a periodic fashion.  In fact, in the absence of spatial derivatives, Eq.\ (\ref{lageq}) is identical to the first model presented in Ref.\ \cite{Shapere:2012nq}.
In their paper, those authors point out 
that the relativistic version of the model suffers from gradient instabilities, that could be cured by
considering instead a Lagrangian of the form 
$+\lambda X -\kappa X^2 + \eta X^3$ whose energy is bounded from below.  However, that was not the kind of model considered in Ref.\ \cite{Mersini-Houghton:2025ybm}.  Here I will explicitly show how the gradient instability in (\ref{lageq}) leads to a loss of energy in the homogeneous mode that is supposed to provide the dark energy, driving it to negative values.

I start with a description of the dynamics of the homogeneous solutions, since we need to perturb around them to exhibit the instability.
The Lagrangian equation of motion is
\be
 \ddot\phi (\kappa -\lambda\dot\phi^2) = {dV\over d\phi}\,,
\label{zeroth}
\ee
which has a first integral of motion, the conserved energy density
\be
    E = V - \sfrac12\dot\phi^2\left(\kappa - \sfrac32\lambda\dot\phi^2\right)\,.
        \label{energy-eq}
\ee
In Ref.\ \cite{Mersini-Houghton:2025ybm}, the form of the potential was not specified (except for being positive semidefinite); I will assume $V = \sfrac12 m^2\phi^2$ for simplicity.  

\begin{figure*}[t]
\centerline{\includegraphics[width=\columnwidth]{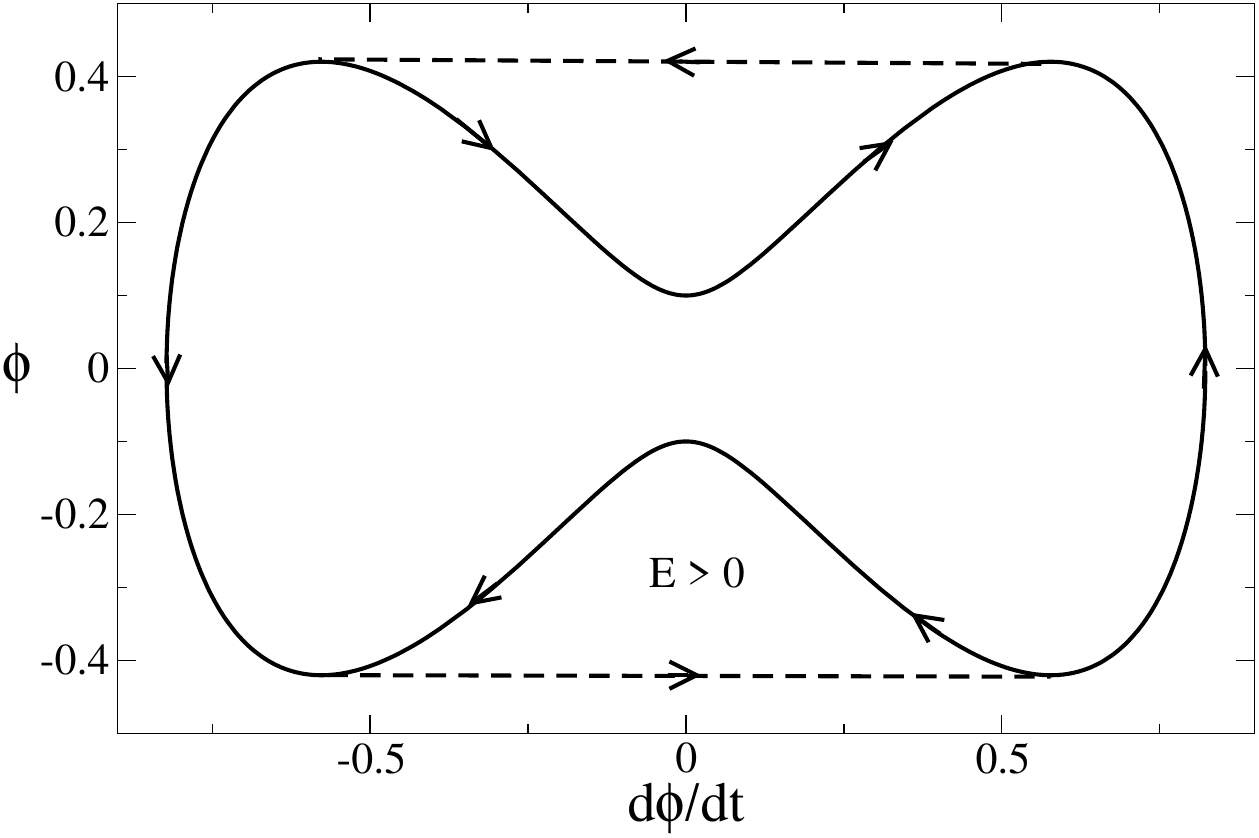}\includegraphics[width=\columnwidth]{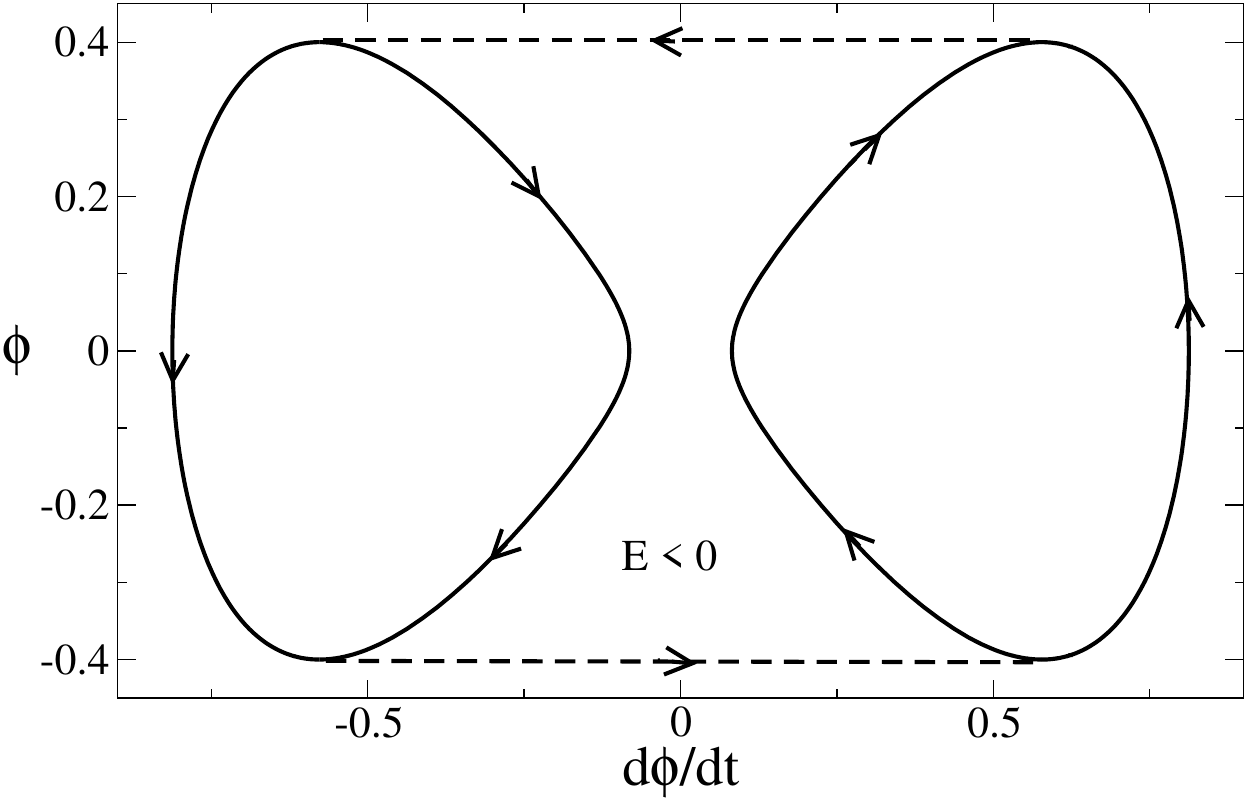}}
\caption{Phase space trajectories for closed orbits in the time crystal model.  The dashed lines indicate where the field velocity discontinuously reverses its sign after hitting the ``brick wall.'' Left plot shows positive energy and right shows negative energy solutions.  In either case there is a bifurcation point after the velocity changes sign, and the field has to ``decide'' which trajectory to follow.  When $E=0$,
an additional bifurcation point appears at the origin.
}
\label{fig:orbit}
\end{figure*}

As Ref.\ \cite{Shapere:2012nq} explained, the dynamics associated with (\ref{energy-eq}) are those of a particle bouncing between two brick walls: the field velocity changes sign suddenly when it reaches a value that minimizes the kinetic energy, $\dot\phi^2 = \kappa/(3\lambda)$.  This can be understood by plotting the phase-space trajectories for fixed energy, as in Fig.\ \ref{fig:orbit}.  A strange feature of these trajectories is the presence of bifurcation points after the velocity jumps, where there is a choice of two possible solutions to join onto.  At these points, $\dot\phi$ can either start decreasing or increasing, while $\phi$ can only either increase or decrease.  The four branches, that are separated by the two bifurcation points plus their reflections, correspond to the four roots of the quartic equation (\ref{energy-eq}) for $\dot\phi$.

Next, we perturb the full equation of motion, including field gradients, around a spatially homogeneous solution $\phi_0(t)$.  The result is
\bea
    (\kappa&-&3\lambda\dot\phi_0^2)\,\delta\ddot\phi -
    (\kappa -\lambda\dot\phi_0^2)\,\nabla^2\delta\phi
\nn\\&+&\left(6\lambda m^2\dot\phi_0 \phi_0\over \kappa-\lambda\dot\phi_0^2\right)\delta\dot\phi
- m^2\delta\phi = 0\,.
\eea
In the damping term $\delta\dot\phi$ we used the zeroth order equation (\ref{zeroth}) to eliminate $\ddot\phi_0$.  Except for $m^2 = V''(\phi)$, all the coefficients are time-dependent since $\phi_0$ is oscillating.  However, we are interested in the secular growth of $\delta\phi$, so we can average these coefficients over one period.  The average value of $\dot\phi_0^2$ is of order $\kappa/(6\lambda)$.  
Moreover we can take $\delta\phi \sim e^{i(\omega t - \vec k\cdot\vec x)}$.  And the coefficient of the damping term averages to zero because $\phi_0\dot\phi_0$ changes sign midway through each 
oscillation.  Then approximately
\be
    -\sfrac12\omega^2 + \sfrac56 k^2 -{m^2\over\kappa} = 0\,.
\ee
This gives imaginary frequencies for wave numbers in the instability band $k\lesssim m/\sqrt{\kappa}$, with
\be
    \omega \sim \pm i\sqrt{2m^2/\kappa -5k^2/3}\,.
\ee

To produce the unstable fluctuations, energy must be
removed from the homogeneous mode, at a rate of order
$m/\sqrt{\kappa}$.  In the present peculiar theory, one should verify that these fluctuations actually carry positive energy in order to be sure
that energy is really lost from the homogeneous mode. For this purpose, we expand the Lagrangian to second order in the perturbation around the homogeneous background.  In the extreme case where $\dot\phi_0^2=
\kappa/(3\lambda)$ which we will argue below is the endpoint of the evolution of the system, this gives
\be
    \delta^2 {\cal L}  = \sfrac12\kappa(\delta\dot\phi)^2
    + \sfrac13\kappa(\nabla\delta\phi)^2 - \sfrac12 m^2(\delta\phi^2)\,.
\ee
Thus the kinetic energy of the fluctuation is positive, but the gradient energy is negative.  Nevertheless, the total energy density of the fluctuations is given by
\be
    E_{\delta\phi} = \left[\kappa(\sfrac12|\omega|^2 - \sfrac13 k^2) +\sfrac12 m^2\right](\delta\phi)^2\,,
\ee    
and one can verify that for $k$ within the instability
band, it is positive.

Notice that $\kappa$ is dimensionless, presumably $O(1)$, although Ref.\ \cite{Mersini-Houghton:2025ybm} gives us no indication as to what values of $\kappa$, $\lambda$, or $m$ might be desirable.   By evaluating the energy density $E$ at the turning point, where $\phi$ attains its maximum amplitude $\phi_t$, we have
\be
    E = \sfrac12 m^2\phi_t^2 - {\kappa^2\over 12\lambda}\,.
\ee

As a consequence of the instability, $\phi_t$ decreases with time at the rate $|\omega|\sim m/\sqrt{\kappa}$, until the homogeneous mode degenerates into a zero-amplitude oscillation that
has infinite frequency, with $\dot\phi_0$ continuously switching between $\pm \sqrt{\kappa/(3\lambda)}$.  The energy density of the homogeneous component reaches a minimum value of $-{\kappa^2/(12\lambda)}$ and the Universe will evolve differently than expected from a typical dark energy model.  In particular, both the energy density and the pressure $p = -5\kappa^2/(36\lambda)$ are negative, which will have peculiar consequences,
perhaps only sensible if the contributions from the 
inhomogeneous components are also taken into account. Preliminary study suggests that the fluctuations diverge and lead to a big rip at late times.
Whether it is possible to choose values of parameters that postpone this until sufficiently far in the future (requiring
$m/\sqrt{\kappa}\lesssim H_0$, the present Hubble rate), while allowing the model to be a viable description of dark energy in the present, would require further study to determine.

{\bf Note added.}  E.\ McDonough pointed out that the stability criteria for  models of the form (\ref{lageq}) are well known in the dark energy literature (see for example Eqs.\ (142,143) of Ref.\ \cite{Copeland:2006wr}): ${\cal L}_{,X} \ge 0$; ${\cal L}_{,X} + 2X {\cal L}_{,XX} \ge 0$.  These are both violated in the present model around the ground state where $X = \kappa/(6\lambda)$.

\medskip
I thank S.\ Caron-Huot, A.\ Frey, L.\ Mersini-Houghton and E.\ McDonough for enlightening discussions.

\bibliographystyle{utphys}
\bibliography{sample}

\end{document}